\documentstyle[aas2pp4]{article}

\begin{document}

\title{Departures From Axisymmetric Morphology and Dynamics in Spiral Galaxies}
\author{David A. Kornreich\altaffilmark{1}}  
\affil{Center for Radiophysics and Space Research} 
\affil{Cornell University Space Sciences Building, Ithaca NY 14853}
\affil{kornreic@astrosun.tn.cornell.edu}

\author{Martha P. Haynes}  
\affil{Center for Radiophysics and Space
Research and National Astronomy and Ionosphere Center\altaffilmark{2}}
\affil{Cornell University Space Sciences Building, Ithaca NY 14853}
\affil{haynes@astrosun.tn.cornell.edu}

\author{R. V. E. Lovelace}
\affil{Cornell University Department of Astronomy}
\affil{Cornell University Space Sciences Building, Ithaca NY 14853}
\affil{rvl1@cornell.edu}

\and

\author{Liese van Zee\altaffilmark{3}}
\affil{National Radio Astronomy Observatory\altaffilmark{4}}
\affil{Liese.vanZee@hia.nrc.ca}

\altaffiltext{1}{NASA Space Grant Graduate Fellow.}
\altaffiltext{2}{The National Astronomy and Ionosphere Center is
operated by Cornell University under a cooperative agreement with the
National Science Foundation.}
\altaffiltext{3}{Present Address: Herzberg Institute of Astrophysics,
5071 W. Saanich Rd, Victoria BC, V8X 4M6, Canada.}
\altaffiltext{4}{The National Radio Astronomy Observatory is a
facility of the National Science Foundation, operated under a
cooperative agreement by Associated Universities, Inc.}

\begin{abstract}
New \ion{H}{1} synthesis data have been obtained for six face--on
galaxies with the Very Large Array. These data and reanalyses of three
additional data sets make up a sample of nine face--on galaxies
analyzed for deviations from axisymmetry in morphology and
dynamics. This sample represents a subsample of galaxies already
analyzed for morphological symmetry properties in the {\it
R-\/}band. Four quantitative measures of dynamical nonaxisymmetry are
compared to one another and to the quantitative measures of
morphological asymmetry in \ion{H}{1} and {\it R-\/}band to
investigate the relationships between nonaxisymmetric morphology and
dynamics. We find no significant relationship between asymmetric
morphology and most of the dynamical measures in our sample. A
possible relationship is found, however, between morphology and
dynamical position angle differences between approaching and receding
sides of the galaxy.
\end{abstract}


\section{Introduction\label{intro}}
Despite the fact that most studies of spiral galaxy dynamics
concentrate on understanding the properties of axisymmetric disks,
evidence is accumulating that many galaxies lack such overall
symmetry. Baldwin {\it et al.\/} (1980)\markcite{blbs} were first to
seriously examine the asymmetries in galactic disks, pointing out that
lopsided galaxies were a common phenomenon not localized to interacting
pairs. 

More recently, frequency of morphological asymmetry in galactic disks
has been quantitatively studied at optical wavelengths. Based on
optical morphology, approximately 30\% of disk galaxies exhibit
significant ``lopsidedness'' (Rix \& Zaritsky 1995\markcite{rz95};
Kornreich {\it et al.\/} 1998\markcite{khl}, hereafter KHL). These
studies are based on data sets containing some 30 targets each.

Dynamical asymmetry, too, has been examined in large samples of
single--dish \ion{H}{1} line profiles. Richter \& Sancisi
(1994)\markcite{rs94}, followed by the 104 galaxy sample of Haynes
{\it et al.\/} (1998, hereafter HHMRvZ)\markcite{h98a}, examined the
symmetry properties of single--dish \ion{H}{1} line profiles and
determined that as many as $\sim50$\% of spiral galaxies show
departures from the expected symmetric two--horned
profile. Nevertheless, asymmetries in line profiles are ambiguous
evidence at best for disturbed dynamics, since they combine both
dynamic and spatial information.

Until recently, however, little work had been done to examine the
symmetry properties of neutral hydrogen in synthesis data. As a
result, the connection between disturbed morphology and disturbed
dynamics in field disk galaxies is only now beginning to be seriously
studied. Schoenmakers {\it et al.\/} (1997, hereafter
SFdZ)\markcite{SFdZ} outline a method for measuring small deviations
from axisymmetry of the potential of a filled gas disk by breaking
down the observed velocity field into its harmonic
components. Recently, Swaters {\it et al.\/} (1999, hereafter
S3vA)\markcite{S3vA} have applied this method to the \ion{H}{1}
synthesis data of two galaxies, DDO~9 and NGC~4395, where the hallmark
of dynamical asymmetry is found to be asymmetry in the rotation curve,
where one side of the curve rises more steeply than the
other. \ion{H}{1} synthesis observations of several other galaxies,
e.g. NGC~3631 (Knapen 1997\markcite{knapen}), NGC~5474 (Rownd {\it et
al.\/} 1994, hereafter RDH\markcite{rownd}), and NGC~7217 (Buta {\it
et al.\/} 1995\markcite{buta}), have also curiously revealed
asymmetries or offsets between the optical centers of light and the
kinematic centers of the neutral gas. While much evidence exists that
many field spirals exhibit non--axisymmetric morphology, and that many
field spirals exhibit asymmetric dynamics, the question of whether the
two phenomena are correlated, or even represent two pictures of a
single underlying effect, remains open.

Both optical morphology and gas dynamics provide clues as to the
overall structure of a galaxy. For instance, Zaritsky \& Rix
(1997)\markcite{zr97} proposed that optical lopsidedness arises from
tidal interactions, minor mergers, or possibly gradual
accretion. Conselice {\it et al.\/} (2000)\markcite{conselice}, on the
other hand, find a correlation between optical asymmetry and \bv\ 
color, and are able to use morphological asymmetry to identify whether
starbursts in a given galaxy are likely caused by interactions and
mergers. Similarly, the neutral hydrogen dynamics of asymmetric
galaxies should be able to distinguish between tidally deformed
galaxies and those which are dynamically isolated. For instance, while
the strongly optically--lopsided galaxy NGC~5474 is well--known to be
under the tidal influence of its neighbor M101, the disturbed
morphologies of the other relatively isolated objects in the KHL
sample are not well--explained by standard tidal interaction models,
which require particular dynamics as well as interactions with
(unobserved) nearby companions.

Alternatively, asymmetries may arise from the excitation of unstable,
one--armed spiral modes possibly triggered by a past interaction or
minor merger (Taga \& Iye 1998a, 1998b; Lovelace {\it et al.\/}
1999)\markcite{love}\markcite{taga1}\markcite{taga2}. Librations of
the optical galaxy about the minimum of the gravitational potential
might also be set in motion by a previous interaction. As density is
dependent on radius in a galaxy, one might expect that the natural
modes of a galaxy also depend on radius. The resulting differential
oscillation might result in both lopsided appearance and kinematic
decoupling from the optical light distribution. An understanding of
these modes, if they exist, could provide a direct measure of the dark
matter distribution (Jog 1997\markcite{jog}; SFdZ\markcite{SFdZ}).

Another type of nonaxisymmetry, warping of galactic disks due to
non-planar motions, has been proposed as an indicator of inclined
flattened halo potentials (e.g. Dekel \& Shlosman 1983\markcite{ds83};
Toomre 1983\markcite{T83}), which are required to stabilize the warp
against differential precession, as well as observational indicators
of massive dark halos (Tubbs \& Sanders 1979)\markcite{sand}. Warps
may also be due to tidal interactions and accretion, and have been
reported to be related to non--circular motions, particularly $m=1$
modes (Weinberg 1998)\markcite{W98}. Although ``sloshing'' librations
in the plane of the galaxy would not imply a correlation between warps
and lopsided dynamics, ``flapping'' librations normal to the galactic
plane could contribute to warping.

In this paper, we present HI synthesis data for nine galaxies whose
optical asymmetry properties were quantified by KHL. Data obtained
from the archives at the Very Large Array (VLA)\footnote{The VLA is a
facility of the National Radio Astronomy Observatory.} have been
reanalyzed for the galaxies NGC~5474, NGC~5701, and UGC~12732, and new
VLA data are presented for the galaxies NGC~991, NGC~1024, UGC~3685,
NGC~3596, UGC~6420, and NGC~4688. These galaxies represent a sample of
face--on galaxies selected on the basis of their optical properties.

In \S \ref{obs}, we discuss the acquisition and reduction of the new
and archived data for our sample of nine galaxies. In
\S\ref{correlations}, we analyze the data obtained for deviations from
morphological and dynamical axisymmetry and describe methods of
determining the magnitudes of such deviations. These quantitative
measures are useful for quantifying warps in face--on galaxies and for
non--circular motions in more inclined galaxies. We then use the
sample to draw correlations between the symmetry parameters of the
galaxies to one another and to global \ion{H}{1} properties. Finally,
in \S\ref{conclusions}, we present our conclusions in the context of other
recent work on galaxy asymmetry.

\section{Observations and Data Reduction\label{obs}}

\subsection{Target Selection and Observations}

The galaxy sample used in this work includes nine galaxies from the
sample of KHL. Table \ref{galsample} summarizes the catalogued data
from the {\it Third Reference Catalogue of Bright Galaxies\/} (de
Vaucouleurs {\it et al.\/} 1991, hereafter RC3)\markcite{rc3} for each
of the targets: column (1) gives the NGC or UGC designations of the
targets; columns (2) and (3) give the right ascension and declination
of the targets in B1950.0 coordinates; column (4) lists the
morphological types; column (5) lists the major and minor blue
diameters, $D_{25}$ and $d_{25}$; and column (6) gives the face--on
blue apparent magnitude corrected for internal and galactic
extinction, $B_T^0$.

\begin{deluxetable}{lrrlrr}
\tablewidth{0pt}
\tablecaption{Summary of Galaxy Sample \label{galsample}}
\tablehead{
\colhead{(1)}&
\colhead{(2)}&
\colhead{(3)}&
\colhead{(4)}&
\colhead{(5)}&
\colhead{(6)}
\\[.2ex]
\colhead{Galaxy}&
\colhead{$\alpha$}&
\colhead{$\delta$}&
\colhead{Type}&
\colhead{$D_{25}\times d_{25}$}&
\colhead{$B_T^0$}
\\[.2ex]
\colhead{}&
\colhead{(B1950)}&
\colhead{(B1950)}&
\colhead{(RC3)}&
\colhead{(arcmin)}&
\colhead{} 
}
\startdata
{\it Original Observations} \cr
NGC 991  & 023302.1 & $-$072234 & SAB(rs)c & $2.7\times 2.4$ & 12.30\cr
NGC 1042 & 023756.3 & $-$083850 & SAB(rs)cd& $4.7\times 3.6$ & 11.38\cr
UGC 3685 & 070433.1 & +614029   & SB(rs)b  & $3.3\times 2.8$ & 12.30\cr
NGC 3596 & 111227.9 & +150338   & SAB(rs)c & $4.0\times 3.8$ & 11.67\cr
UGC 6429 & 112225.1 & +640011   & SA(rs)c  & $2.1\times 1.7$ & 13.63\cr
NGC 4688 & 124513.8 & +043558   & SB(s)cd  & $3.2\times 2.8$ & 12.47\cr
{\it Archive Data} \cr
NGC 5474 & 140314.6 & +535403   & SA(s)cd Pec&$4.8\times 4.3$ & 11.27\cr
NGC 5701 & 143641.5 & +053450   & (R)SB(rs)0/a&$4.3\times 4.1$& 11.69\cr
UGC 12732& 233808.8 & +255727   & Sm       & $3.0\times 2.8$  & 13.55\cr
\enddata
\end{deluxetable}

\subsubsection{Original Data Acquisition}

Of the optical sample in KHL, the six galaxies NGC~991, NGC~1042,
UGC~3685, NGC~3596, UGC~6420, and NGC~4688 were selected for
\ion{H}{1} 21-cm synthesis observations based on their apparent degree
of asymmetry and position in the sky such that they could be observed
at night during the time scheduled for the CS array.  As the
photometric analysis conducted in KHL had not yet been completed at
the time of the target selection, estimates of asymmetry were made by
eye. Night--time observations were preferred in order to minimize
solar interference. The KHL sample was an optically selected sample of
disk galaxies chosen on the basis of face--on appearance to permit
estimation of the optical morphological asymmetry unencumbered by
factors such as absorption by dust and inclination errors. The KHL
selection criteria were: small axial ratio, narrow (typically $W_{50}
\lesssim 100$~km~s$^{-1}$) velocity width obtained from the private
database of Giovanelli \& Haynes known as the Arecibo General Catalog
(AGC), and (except for NGC~5474 and NGC~1042) isolation from obvious
companions. See KHL for details of the optical target selection.

\ion{H}{1} observations of these targets were conducted in August
1997, December 1998, and January 1999 with the VLA in its CS
configuration. The CS configuration is a modification of the C
configuration which provides the resolution of the C array while
maintaining a sufficient number of short spacings to provide sampling
of large--scale structure by positioning one or more antennas at D
configuration locations. In 1997, the CS configuration consisted of
the standard C array with two antennae (from the middle of the east
and west arms) relocated at inner D configuration stations.  In 1998
and 1999, one antenna was taken from the middle of the north arm and
placed at an inner D configuration station to provide the requisite
short baselines.

Each galaxy was observed for a total of approximately 5.5 hours on--source
integration time in 2AD correlator mode with 127 channels of 12.207
kHz separation after Hanning smoothing. This experimental setup
provided for a total bandwidth of 1.5625 MHz and, at 1.4 GHz, for 2.6
km~s$^{-1}$ resolution. Observations were flux and bandpass calibrated by
observing flux calibrators for 15 minutes immediately before and after
the target observations, and phase calibrators for 5 minutes at 30
minute intervals during the target observations.

\subsubsection{VLA Archival Data\label{betterdata}}

In addition to the data obtained by us at the VLA, several data sets
obtained by other observers were recovered from the VLA data archives
for reanalysis. These data sets include observations of the galaxies
NGC~5474 (\markcite{rownd}RDH), NGC~5701 (observed by S. E. Schneider
in 1986), and UGC~12732 (\markcite{schul}Schulman et al. 1997). These
targets were chosen because each was analyzed for optical
morphological asymmetry as part of the KHL sample, and VLA archive
\ion{H}{1} line data of sufficient sensitivity (about
$10^{20}$~cm$^{-2}$) and spectral resolution (less than
15~km~s$^{-1}$) for the current analysis existed for them.

The archival data were retrieved as raw {\it UV\/} visibility data,
and reduced by the same methods as were used for our original
observations. The raw data sets for NGC~5701 and UGC~12732 were
Hanning smoothed on--line by their observers; that of NGC~5474 was not.

An observational summary of the nine VLA data sets in the total sample
is presented in Table \ref{archtab}. Columns in this table represent:
(1)~The NGC or UGC designation of the target; (2)~The VLA
configuration in which the data were obtained; (3)~The date of the
observation; (4)~The observer or paper in which the observation was
reported; (5)~The total on--source integration time, in minutes;
(6)~The total bandwidth in MHz; (7)~The channel separation in
km~s$^{-1}$; (8)~The total number of channels in the data set; (9)~The
synthesized clean beam size in our natural weighted maps, in
arcseconds; and (10)~The RMS noise $\sigma$ observed in the line--free
channels, in mJy~beam$^{-1}$.

\begin{deluxetable}{lcrllrrrrc}
\tablenum{2}
\tablewidth{0pt}
\tablecaption{Observational Summary \label{archtab}}
\tablehead{
\colhead{(1)}&
\colhead{(2)}&
\colhead{(3)}&
\colhead{(4)}&
\colhead{(5)}&
\colhead{(6)}&
\colhead{(7)}&
\colhead{(8)}&
\colhead{(9)}&
\colhead{(10)}
\\[.2ex]\colhead{Galaxy}&
\colhead{Array}&
\colhead{Date Obs.}&
\colhead{Reference}&
\colhead{$t_{int}$}&
\colhead{$W_{band}$}&
\colhead{$W_{chan}$}&
\colhead{$N_{chan}$}&
\colhead{Beam}&
\colhead{$\sigma$}
\\[.2ex]
\colhead{}&
\colhead{}&
\colhead{}&
\colhead{}&
\colhead{(min)}&
\colhead{(MHz)}&
\colhead{(km s$^{-1}$)}&
\colhead{}&
\colhead{(arcsec)}&
\colhead{(mJy beam$^{-1}$)}
}
\startdata
NGC 991  & CS & Aug 97 & This Work         & 381   & 1.5625& 2.60 & 127& $25\times 20$&0.8\cr
NGC 1042 & CS & Aug 97 & This Work         & 380   & 1.5625& 2.60 & 127& $23\times 18$&0.8\cr
UGC 3685 & CS & Aug 97 & This Work         & 380.5 & 1.5625& 2.60 & 127& $19\times 18$&0.7\cr
NGC 3596 & CS & Jan 99  & This Work         & 503.5 & 1.5625& 2.60 & 127& $19\times 18$&0.7\cr
UGC 6429 & CS & Jan 99  & This Work         & 377 & 1.5625  & 2.64 & 127& $20\times 16$&0.6\cr
NGC 4688 & CS & Dec 98  & This Work         & 370.5 & 1.5625& 2.59 & 127& $21\times 18$&0.8\cr
NGC 5474 & D & Jul 83  & Rownd {\it et al.} (1994)     & 287.5 & 3.125  & 2.58 & 31 & $54\times 51$&2.2\cr
NGC 5701 & D & Apr 87  & S. E. Schneider&750.5&3.125& 10.41& 31 & $70\times 63$&1.3\cr
UGC 12732& D & Aug 92  & Schulman {\it et al.} (1997)  & 660   & 3.125  & 10.36& 63 & $54\times 53$&0.4\cr
\enddata
\end{deluxetable}

The archival data sets were obtained between early 1983 and 1993, with
the earlier data sets consisting of a small number of channels in the
D array. While we expected to improve marginally upon the previous
data reduction for these archival data sets, we were surprised to find
that due to the exponential improvement in computer technology since
that time, the reanalysis of these data yielded a remarkable
improvement over the initial analyses. By cleaning the data to deeper
levels than was possible in the past, we were able to measure
\ion{H}{1} column densities up to almost an order of magnitude fainter
than those previously reported for the same data sets. As a specific
example, we were able to iteratively clean the {\it UV\/} data for
NGC~5474 for 200,000 iterations, down to the $1\sigma$ theoretical
noise level, as compared to the 500 iterations reported by RDH. Our
reanalysis was thus able to trace the \ion{H}{1} to a $1\sigma$ column
density of $6\times 10^{18}\ {\rm cm}^{-2}\ {\rm chan}^{-1}$, compared
to the $1.2\times 10^{19}\ {\rm cm}^{-2}\ {\rm chan}^{-1}$ quoted in
RDH, an improvement of about half an order of magnitude with the same
data set. This greater sensitivity allowed us to uncover the very
peculiar outer dynamics of this galaxy not observed by RDH, structure
which is of dynamical importance. As expected, such large gains in
sensitivity were most pronounced from the earlier data sets, and of
less importance for later ones.

\subsection{UV Processing and Data Analysis}

Reduction of the visibility data was conducted using the Astronomical
Image Processing System (AIPS)\footnote{AIPS is distributed by the
National Radio Astronomy Observatory.} (Napier {\it et al.\/}
1983)\markcite{aips}. Data were examined via the program TVFLG, in
which poor quality {\it UV\/} data were identified and
flagged. Following the flagging procedure, a linear interpolation of
bandpass calibrations obtained from the phase calibrators for each
galaxy was obtained and applied to the target data. 

The continuum was then subtracted from each data set using the program
UVLIN (Cornwell {\it et al.\/} 1992)\markcite{uvlin}, which subtracts
continuum emission from all channels based on a linear fit of the
visibilities in a given number of line--free channels, as selected by
manual inspection. The total number of line--free channels used varied
according to the data set, but typically numbered about 30 for
originally acquired data, and 10 for archive data. The {\it UV\/} data
were then transformed to the XY plane using the program IMAGR.

At this point in the calibration, it was discovered that the
catalogued declination of the phase calibrator used with the UGC~6429
data set was incorrect by one arcminute. This had the effect of
producing incorrect coordinates in the XY data cubes. This problem was 
solved by identifying continuum sources observed in the raw image
with sources catalogued in the NRAO VLA Sky Survey (Condon {\it et
al.\/} 1998)\markcite{nvss} and performing astrometry on these sources.

For each galaxy, data cubes were constructed using both natural and
robust weighting. We present in Figure \ref{chanmaps} the data cubes
in the form of channel maps derived by natural weighting, because
natural weighting appeared to strike the best balance between the
signal--to--noise ratio and resolution. This figure presents the
channels containing signal for each galaxy, bracketed by several noise
channels. The channel velocity for each subframe is given in the upper
left corner. For most of the targets, which had emission in a large
number of channels, only every other channel is depicted. For UGC~6429
and NGC~4688, which were detected in a smaller number of channels,
every channel containing signal is shown.

Following this standard visibility reduction, the data image cubes
were transferred for further reduction in the GIPSY environment (van
der Hulst {\it et al.\/} 1992\markcite{gipsy}, {\it GIPSY\/}
2000\markcite{gip2000}). Copies of the data cubes were then smoothed
to provide easier separation of regions with signal from regions with
no signal. Data in these copies were cropped at the $3\sigma$
level. The cropped regions were then mapped back to the unsmoothed
cubes, which then consisted only of regions containing
signal. Following correction for primary beam, these regions were
integrated along the velocity axis into zeroth, first, and second
moments of the flux, thus creating three spatial maps, representing
the \ion{H}{1} flux, bulk motion, and velocity dispersion
respectively. To eliminate spurious noise peaks during this
integration, a ``windowing'' function was used, such that only peaks
seen in two or more consecutive channels were included in the
integration. The \ion{H}{1} flux was then scaled to the column density
$N$ of the gas, assuming optically thin \ion{H}{1}, via the relation
\begin{equation}
N = 1.10297\times 10^{24}\;{\rm cm}^{-2} \int \frac{S}{ab} \left(1+v/c\right)^2\,dv
\end{equation}
where $S$ is the total flux density in Jy/Beam, and $a$ and $b$ are
the major and minor FWHM axes of the Gaussian beam in arcseconds.

Following construction of the moment maps, simulated single--dish line
profiles for each galaxy were obtained by integrating the data cubes
over the spatial axes. Total integrated flux measurements were
obtained by integrating the line profiles along the velocity
axis. They were then compared to single--dish data obtained from
various sources. These single--dish data were corrected for pointing
errors and beam smearing following the two--Gaussian distribution
model of Hewitt {\it et al.\/} (1983)\markcite{HHG}, with the Gaussian
parameters derived from the synthesis data obtained here by fitting a
double Gaussian to the moment zero maps. The comparison of synthesis
to corrected single--dish flux is presented as part of Table
\ref{globalHI}. Following the discussion of Hewitt {\it et al.\/}, we
expect between 15\% and 20\% uncertainty in the derived corrected
flux. In general, there is agreement between the total flux
measurements derived here and the single-dish data within the errors.

\markcite{FT}\markcite{SS}
\begin{deluxetable}{lrrrrrrrrrrr}
\tablenum{3} \label{globalHI}
\tablewidth{0pt}
\tablecaption{Global \ion{H}{1} Properties}
\tablehead{
\colhead{(1)}&
\colhead{(2)}&
\colhead{(3)}&
\colhead{(4)}&
\colhead{(5)}&
\colhead{(6)}&
\colhead{(7)}&
\colhead{(8)}&
\colhead{(9)}&
\colhead{(10)}
\\[.2ex]\colhead{Galaxy}&
\colhead{$F_{\rm VLA}$}&
\colhead{$F_{\rm SD}$}&
\colhead{$W_{20}$}&
\colhead{$V_{HI}$}&
\colhead{$D_{HI}/D_{25}$}&
\colhead{$V_{rot}\sin i$}&
\colhead{Dist}&
\colhead{$M_{HI}$}&
\colhead{$M_{HI}/L_B$}
\\[.2ex]
\colhead{}&
\multicolumn{2}{c}{(Jy km s$^{-1}$)}&
\multicolumn{2}{c}{(km s$^{-1}$)}&
\colhead{}&
\colhead{(km s$^{-1}$)}&
\colhead{(Mpc)}&
\colhead{($10^9 M_\odot$)}&
\colhead{($M_\odot / L_\odot$)}
}
\startdata
NGC 991  & 21 & 21\tablenotemark{a} & 91  & 1531 & 1.9 & 33.0 &20       & 2.0 &  .27\cr
NGC 1042 & 55 & 56\tablenotemark{a} & 116 & 1372 & 1.7 & 43.7 &20       & 5.2 &  .30\cr
UGC 3685 & 43 & 55\tablenotemark{a} & 97  & 1796 & 2.1 & 36.8 &30       & 9.1 &  .54\cr
NGC 3596 & 29 & 33\tablenotemark{b} & 127 & 1193 & 1.1 & 51.7 &14       & 1.3 &  .20\cr
UGC 6429 & 10 & 11\tablenotemark{c} & 69  & 3727 & 1.5 & 23.3 &75       & 13  &  .42\cr
NGC 4688 & 28 & 35\tablenotemark{a} & 60  &  984 & 1.8 & 18.1 &17       & 1.9 &  .41\cr 
NGC 5474 & 78 & 120\tablenotemark{a}& 60  &  277 & 1.9 &  8.2 & 7       & 0.9 &  .40\cr
NGC 5701 & 58 & 63\tablenotemark{b} & 139 & 1509 & 2.6 & 57.1 &26       & 9.3 &  .42\cr
UGC 12732& 79 & 79\tablenotemark{a} & 139 &  746 & 3.1 & 54.6 &15       & 4.2 &  3.2\cr
\enddata
\tablenotetext{a}{Fisher \& Tully 1981}
\tablenotetext{b}{HHMRvZ}
\tablenotetext{c}{Staveley--Smith \& Davies 1988}
\end{deluxetable}

The first order moment maps were used to obtain a kinematic model of
the galaxies by iteratively applying the standard GIPSY routine ROTCUR
(Begeman 1989)\markcite{rotcur}. This routine divides the galaxy into
a number of rings, and fits a function
\begin{equation}
V(x,y) = V_{sys} + V_{rot} \cos\theta \sin i
\end{equation}
where
\begin{equation}
\cos\theta = \frac{-\left(x-x_0\right) \sin(\Gamma) + \left(y-y_0\right) \cos(\Gamma)}{r}
\end{equation}
to the velocity field in each ring, allowing the systematic velocity
$V_{sys}$, the rotation velocity $V_{rot}$, inclination $i$, position
angle $\Gamma$, and kinematic center $\left(x_0,y_0\right)$ to
vary. 

Because of the face--on aspect of the target galaxies, adequate
kinematic solutions could be found, but were spread over a large area
of parameter space. Specifically, although $V_{rot}\sin i$ remained
well--constrained for each ring, changes in $V_{rot}$ and $i$ were
found to be interchangeable with little effect on the quality of fit
in the model function. The result was often that the algorithm
produced unphysical values of $V_{rot}$ with wildly varying
inclinations if all parameters were allowed to vary independently. For
this reason, the ROTCUR model was applied iteratively to the data,
alternately holding $V_{rot}(r)$, $i(r)$, and $(x_0,y_0)$ fixed. We
present rotation curves and all subsequent analysis in terms of
$V_{rot}\sin i$. We caution, therefore, that variations in the
$V_{rot}\sin i (r)$ curves we present may represent either actual
changes in the rotation velocities of the galaxies, changes in the
galaxies' inclinations, or a combination of both. Inability to further
constrain $V_{rot}(r)$ also precluded the construction of reliable
model velocity fields which could reasonably be expected to reproduce
the potentials of the galaxies. Once the tilted ring fit was applied
iteratively to the entire galaxy to obtain initial values of the
parameters, the model was fitted to both halves of the galaxy
separately, so that $V_{rot}\sin i (r)$ and position angles could be
determined for both sides independently.

Position--velocity diagrams for one beam--width along the major axis
of each data set were then obtained.  To construct the
position--velocity diagrams, slices of the data cubes were taken
through the kinematic centers obtained in the model fitting. Each
slice was a strip of data whose spatial length was the total detected
extent of the galaxy, spatial width was the width of a beam, and
frequency depth was the depth of the cube. The spatial position angle
of the slice was determined by averaging the model position angles
over radii $r \leq R_{25}$.  Data were then averaged over the width of
the slice to produce the position--velocity diagrams.

Distances to most of the sample galaxies were determined using a
nonlinear Virgocentric infall model with an infall flow velocity of
the Local Group of 230~km~s$^{-1}$, Hubble parameter $H_0 =
74$~km~s$^{-1}$~Mpc$^{-1}$, distance to Virgo of 16.6 Mpc, and the
mean overdensity of the Virgo cluster compared to the ambient density
of $\rho_{\rm Virgo}/\rho_0=2$. Our model uses these parameters to
solve for the velocity field of the local universe, then to find the
possible distances of individual galaxies by finding the roots of the
difference of the Local~Group--ocentric velocity (derived from the
heliocentric velocity by the method of Mould, Aaronson, and Huchra
(1980)\markcite{LGcentric}) and the velocity found for a given
distance.  This is done by using the {\it Numerical Recipes} (Press
{\it et al.\/} 1986\markcite{NRF}) routine ZBRAK to determine the
bounds of the regions surrounding each root, which are then used to
solve for the roots themselves.

This model yields multiple distances for galaxies in the
triple--valued--region close to the Virgo cluster. Distances of 12,
14, and 24 Mpc were obtained for NGC~3596, and of 7, 17, and 22~Mpc
for NGC~4688. Here, we have taken the value of 14~Mpc for NGC~3596,
based on the distance modulus of 31.41 found by Bottinelli {\it et
al.\/} (1985)\markcite{b85} using the Tully--Fisher relation of
\ion{H}{1} line width and {\it B\/}--band luminosity. We have taken a
distance of 17~Mpc for NGC~4688 based on a study of supernovae by
Gaskell (1992)\markcite{g92} in which a distance of 17~Mpc yields a
peak luminosity of SN~1966B commensurate with other type II--L
supernovae. Following RDH, the distance to NGC~5474 of 7~Mpc was taken
directly from Sandage \& Tammann (1974)\markcite{ST5474}, where
distances to galaxies in the M101 system were found by measuring the
angular sizes of \ion{H}{2} regions. We have also adopted a value of
20~Mpc for NGC~991, equal to that of NGC~1042, because both galaxies
are found by Garcia (1993)\markcite{garcia} to be members LGG~71, a
loose group of galaxies which includes the NGC~1052 group.

The global \ion{H}{1} properties derived from the analysis are
presented in Table \ref{globalHI}. Table columns represent: (1)~The
NGC or UGC designation of the target; (2)~The integrated total flux
observed at the VLA, $F_{\rm VLA}$, in Jy~km~s$^{-1}$; (3)~The single
dish flux, $F_{\rm SD}$, in Jy~km~s$^{-1}$, corrected for pointing
errors and beam dilution as described earlier in this section with
associated reference for the uncorrected flux measurement; (4)~The
full width of the \ion{H}{1} line profile measured at 20\% of the peak
flux, $W_{20}$, in km~s$^{-1}$; (5)~The systematic velocity $V_{HI}$
of the target, in km~s$^{-1}$, obtained from tilted ring modeling of
the data, with typical errors of $\pm 5$~km~s$^{-1}$; (6)~The ratio
of the \ion{H}{1} to optical diameters, $D_{HI}/D_{25}$, where
$D_{HI}$ is measured at a surface density of $1 M_\odot$~pc$^{-2}$.
(7)~The observed value of $V_{rot}\sin i$ averaged over the flat part
of the curve; (8)~The distance to the target, in Mpc, obtained from
the Virgocentric infall model; (9)~Total \ion{H}{1} mass observed at
the VLA, obtained from the flux density via the relation
\begin{equation}
M_{HI}=2.36\times 10^5M_\odot \left(\frac{D^2}{\rm Mpc^2} \right)
\int \frac{S_\nu}{\rm Jy}\,\frac{dv}{\rm km\;s^{-1}},
\end{equation}
where $D$ is the distance to the galaxy and $S_\nu$ is the \ion{H}{1}
flux density in Janskys; and (10) the \ion{H}{1} mass to blue
luminosity ratio $M_{HI}/L_B$, where the luminosity is derived from
$B_T^0$.

\subsection{Qualitative Features of the Target Objects}

General comments regarding the gas morphology and dynamics of each
target are discussed in this subsection. The text describing each
galaxy is accompanied by a figure presenting, in the top two frames,
the \ion{H}{1} column density contours overlaid on the optical image
of the galaxy obtained by KHL, and a grayscale map of the
\ion{H}{1} column density itself. Noise in the column density maps
differs from pixel to pixel due to the windowing function, and was
corrected for noise correlations among Hanning smoothed channels
following the discussion of Verheijen (1997)\markcite{v97}. The noise
in a given pixel is
\begin{equation}
\sigma^l = \frac{4}{\sqrt{6}} \left[(N-3/4) + N^2{\cal N}^2\right]^{\frac12} \sigma^h
\end{equation}
where $N$ is the number of adjacent Hanning smoothed channels which
contributed to the pixel, $\sigma^h$ is the noise in a single channel,
and
\begin{equation}
{\cal N} \equiv \sqrt{\frac{\left(N_1-\frac34\right)}{4N_1^2} +
\frac{\left(N_2-\frac34\right)}{4N_2^2}}
\end{equation}
is a factor representing the noise contributed by subtracting the
first $N_1$ and last $N_2$ line-free channels from the cube. The
average pixel value for all pixels with signal--to--noise of $2.75 <
S/N < 3.25$ was taken to define the $3\sigma$ level for the
map. Column density contours in the maps are selected for each galaxy
individually based primarily on clarity of presentation, and the
\ion{H}{1} grayscales are continuous and linear.

In order to produce the optical overlays, astrometry was performed for
each of the {\it R\/}--band images obtained in KHL corresponding to
the VLA target galaxies, based on star positions obtained from the
{\it HST Guide Star Catalog\/} (Lasker {\it et al.\/}
1992)\markcite{gsc}. This also allowed morphological asymmetry
measurements to be made with respect to the position of the galaxies'
optical nuclei.

The center left frame exhibits the position--velocity diagram along
the \ion{H}{1} dynamical major axis, as determined by averaging over
the position angles for all radii inside of $R_{25}$. The position
angle is given at the top of the plot, and the position axis is
measured in arcminutes from the dynamical center. Dotted vertical
lines in the position--velocity diagram indicate $R_{25}$. Graininess of
the archive data is due to large channel widths and small number of
channels in these sets, which make the corresponding diagrams data--sparse.

The center left frame is a grayscale representation of the \ion{H}{1}
velocity distribution, with grayscale steps equal to the
contours. Again, the contours and grayscales are chosen individually
for each galaxy based on clarity of presentation. Dark grayscales
represent the receding side of the galaxy; lighter grayscales the
approaching side.  Values for selected contours are presented in
km~s$^{-1}$. Overlaid on the grayscale moment frames are circles of
radius $R_{25}$ centered on the center of light of the optical
galaxies, except for NGC~5474, where the center of light is so
displaced from both the dynamical center and the center of the outer
optical isophotes that the dynamical center is used. We have elected
to show circles here, although the RC3 indicates some elongation, as
evident from Table \ref{galsample}, because of ambiguity in extracting
asymmetry versus inclination for face--on objects.

Finally, the lower two frames present the integrated \ion{H}{1} line
profile and $V_{rot}\sin i (r)$ for the galaxy. This latter diagram
presents the data for the line--of--sight rotation velocity
$V_{rot}\sin i$ (km~s$^{-1}$), and position angle (degrees east of
north) from the tilted ring fitting algorithm as a function of radius
in arcseconds. The error values given are the numerical errors in the
fitting algorithm; we estimate that systematic errors are on the order
of twice as large. The data for the approaching side (circles; solid
lines) and the receding side (triangles; dashed lines) are presented
superposed. It is important to note that the ordinate of the rotation
plot does not start at zero; because of beam smearing, we cannot trace
the rotation close to the center. The differences between values for
the approaching and receding sides would be washed out with an
increased scale.

\subsubsection{NGC 991}

NGC~991 (Figure \ref{n991}) is a flocculent SBc galaxy whose
northwestern spiral arms extend twice as far from the nucleus as do
the visible southeastern arms. Since the optical flux of the
flocculent structures northwest of the nucleus is greater than that to
the southeast, it is classified as asymmetric by KHL. The bar is
elongated in the east--west direction. The detected neutral hydrogen
extends to 1.5 times the optical radius. A neutral hydrogen disk of $1.2
\times 10^{21}$~cm$^{-2}$ coexists with the optical disk, except for a
gap over the underluminous southern disk which has an average
\ion{H}{1} column density of $8.5\times 10^{20}$~cm$^{-2}$. This gap
appears in a partial \ion{H}{1} ring with two flux maxima at a radius
of 1\arcmin\ (6.4~kpc). The outer \ion{H}{1} disk is morphologically
symmetric. According to Garcia (1993)\markcite{garcia}, the nearest
neighbor to NGC~991 in LGG~71 is NGC~1022, 34\arcmin\ (198~kpc at the
adopted distance of 20~Mpc) to the northeast.

The \ion{H}{1} dynamics of NGC~991 are relatively undisturbed within
the optical radius as illustrated in the position--velocity diagram
(see Figure \ref{n991}), but certain nonaxisymmetric dynamical
features are evident in the outer disk.  Closed contours on the east
side of the major axis indicate a possible warp in the disk. Strong
twisting of the minor axis velocity contours analogous to the ``S--''
shaped velocity contours of NGC~5474 (see \S\ref{N5474}) is also
present in NGC~991. The same phenomenon can also be seen in the
channel maps as an area of enhanced flux to the east of the galaxy at
1513.8 and 1508.6 km~s$^{-1}$. This is indicative of either a severely
warped disk or strong streaming motions through an outer spiral
arm. The dynamical center of the velocity distribution is consistent
with the {\it R\/}--band center of light of the galaxy, but the
dynamics exhibit a $40^\circ$ maximum shift in position angle
from approaching to receding side for the same radius, as a result of
the detected ``S--'' shaped velocity contours. The synthesised line
profile is asymmetric, with decreasing flux towards the approaching
side.

\subsubsection{NGC 1042 \label{n1042}}

NGC~1042 was first mapped in \ion{H}{1} by van Gorkom {\it et al.\/}
(1986)\markcite{vG86} whose primary purpose was to obtain \ion{H}{1}
synthesis data for the elliptical NGC~1052 and its companions, which
include NGC~1042, some 15\arcmin\ (87~kpc at the adopted distance of
20~Mpc) southwest of NGC~1052.  They point out that the neutral gas in
NGC~1042 is asymmetric in the direction of NGC~1052, suggesting a
tidal interaction. Because these data were collected with only 18
antennae yielding a synthesised beam of $60^{\prime\prime}\times
60^{\prime\prime}$ at a velocity resolution (41~km~s$^{-1}$) more
suited to the inclined galaxy NGC~1052, it was decided that newer data
were required to better constrain the dynamical parameters. These new
data we present here.

NGC~1042 (Figure \ref{n1042fig}) is a nearby asymmetric spiral galaxy
whose eastern spiral arm contains many active \ion{H}{2} regions, but
whose western spiral arm is underluminous. Although the total
\ion{H}{1} flux agrees with previous observations, the \ion{H}{1} mass
derived from our data is larger by a factor of two than that reported
by van Gorkom {\it et al.\/} and references therein, due to
the difference in adopted distances.

The \ion{H}{1} morphology of this galaxy is unremarkable inside the
optical radius, with the \ion{H}{1} column density following the
pattern of the dominant spiral arms outside of the central \ion{H}{1}
hole. Asymmetry in the outer regions, however, is clear. The column
density drop--off towards the northeast is far more gradual than the
precipitous drop--off to the southwest, giving the galaxy the
appearance of compression on the southwestern ``forward'' edge, as if
traveling parallel to its disk through an intergalactic
medium. Interestingly, this would indicate that NGC~1042 is traveling
in the direction opposite to the position of NGC~1052. Van Gorkom {\it
et al.\/} also comment that the \ion{H}{1} in NGC~1042 is not
significantly larger in extent than the optical galaxy. Our estimate
of $D_{HI}/D_{25} = 1.7$, however, is indicative of a large
\ion{H}{1} disk and is close to the average for our sample.

The dynamics of NGC~1042 include substantial evidence along the minor
axis of a warp in the disk, especially north of the optical
disk. Several other nonaxisymmetric kinematic features are indicated
as well. One of these is the closed velocity contours on the eastern
side of the galaxy, seen in both the velocity field and the channel
maps, implying a peak in $V_{rot}$ just inside the optical
radius. This feature is also consistent with a warp. The warped disk
is also manifest in the \ion{H}{1} position angle curve, in which
$\Gamma$ changes by as much as $43^\circ$ as a function of
radius. Streaming motions are apparent on both sides of the minor
axis, correlating with the outer extensions of the two dominant
optical spiral arms. $V_{rot}\sin i (r)$ in NGC~1042 is reminiscent of
rotation curves presented by S3vA, in which opposite sides rise with
different slopes. These dynamical perturbations are suggestive of
tidal interaction with NGC~1052.

An interesting feature to note is that the position angle of the
optical galaxy reported by KHL, found by fitting elliptical isophotes,
appears perpendicular to the major axis of the velocity field of the
\ion{H}{1}. This effect is apparent in Figure \ref{n1042fig}, where
the morphological major axis of the low--surface brightness optical
galaxy appears to run north--south, while the dynamical major axis in
the \ion{H}{1} velocity field is clearly east--west. The implication
is that the optical disk possesses inherent ellipticity, again
suggesting tidal interaction with NGC~1052. Since KHL estimated
inclinations based on optical axial ratios and the assumption of
circular disks, the optical asymmetry parameter derived there is
likely to be an underestimate. In this work, we use the dynamical
position angle. 

\subsubsection{UGC 3685}

UGC~3685 (Figure \ref{u3685}) is a flocculent SBb galaxy whose bar
terminates at a well--defined circular ring. Two major spiral arms
extend outwards from the ring, but their positions are uncorrelated
with the bar. The northern spiral arm and associated flocculent
structure is more luminous than the southern structure, and contains
many bright \ion{H}{2} regions. An additional area of low surface
brightness extends to the northeast. It is classified as symmetric by
KHL. There are no galaxies within 1$^\circ$ and 1000~km~s$^{-1}$ of
UGC~3685 listed in the AGC.

The neutral hydrogen distribution extends to a radius of 2.1 $R_{25}$.
Not all of the single--dish flux was recovered during these
observations, and although comparison with an HI line profile obtained
with the former 91-m telescope at Green Bank available in the digital
archive of \ion{H}{1} spectra of Giovanelli \& Haynes indicates that
much of the difference can be explained by calibration errors,
additional low--surface--density diffuse gas whose flux was not
detected at the VLA cannot be ruled out. The ``fishtail,'' a small
western extension of the distribution, occurs at the
$10^{20}$~cm$^{-2}$ level, and along with the \ion{H}{1} gap to the
northwest and strong spiral structure, significantly affects the
outcome of the morphological analysis. The overall dynamics of the
neutral hydrogen seem relatively undisturbed, although streaming
motions of 3--5~km~s$^{-1}$ are apparent just outside the optical
radius to the north, and at $1.5R_{25}$ to the south, following the
spiral arms. The rotation curve and line profile of the galaxy,
however, are asymmetric between approaching and receding sides.

The most interesting dynamical feature of UGC~3685 appears in the
position--velocity diagram and the channel maps. A region of high
(40~km~s$^{-1}$) velocity dispersion is detected 2$\arcmin$
northwest of the galactic nucleus at an azimuth of approximately
$299\arcdeg$. This feature is consistent with the presence of a
superbubble, an outflow of hot gas into the halo as a result of rapid
star formation and supernovae type II in large \ion{H}{2}
regions. The symmetry in the velocity contours above and below the
systematic rotational velocity of the remainder of the disk implies
that the bubble originated with an event near the midplane of the
disk. A similar \ion{H}{1} bubble has been observed in M101 by
Kamphuis {\it et al.\/} (1991)\markcite{M101bub}. The bubble in
UGC~3685 is observed nearly normal to the disk, allowing the enclosed
kinetic energy of expansion into the halo to be directly measured. The
\ion{H}{1} mass enclosed in the bubble is $1.22\times 10^7 M_\odot$,
resulting in an approximate kinetic energy of $2\times 10^{53}$~ergs,
of the same order as energies found in superbubbles in M101.

\subsubsection{NGC 3596}

NGC~3596 (Figure \ref{n3596}) is an optically asymmetric Sc galaxy
which exhibits two ``grand design'' spiral arms surrounded by a low
surface brightness system of flocculent structure. The contrast of the
optical image in the accompanying figure was selected to emphasize the
flocculent outer structure so that the relationship between this
structure and the \ion{H}{1} morphology is clearer. The neutral gas in
NGC~3596 is distributed in a ring comparable in radius to the bright
spiral arms, from which extends high--column density spiral structure
in the south, while a single \ion{H}{1} spiral arm extends to the
north. The northern arm and the highest column density southern arm
correspond to very low--surface brightness features in the optical;
however, the optical features do not extend coherently along the
entire length of the gas features. While the northern \ion{H}{1} arm
overlies its corresponding flocculent optical arm, the optical feature
associated with the greatest \ion{H}{1} column density in the southern
arm is offset from the \ion{H}{1} by $10^{\prime\prime}$ (0.7 kpc) to
the east. The remaining \ion{H}{1} spiral arms overlie corresponding
low--surface brightness flocculent structures. The overall asymmetric
distribution of neutral gas contributes to a relatively high value of
the morphological asymmetry parameters for this target.

Although there are no galaxies within 1$^\circ$ of NGC~3596 with known
redshifts near 1193~km~s$^{-1}$ listed in the AGC, NGC~3596 is dynamically
perturbed. Although the \ion{H}{1} disk does not extend far beyond the
optical radius, and the inner velocity contours of the galaxy appear
very regular, warping is evident beyond $R_{25}$, and the dynamical
position angle of the approaching side varies southward by as much as
31 degrees. The severity of the warp is most evident in the channel
maps, in which the disk appears so warped that several channels on
both the approaching and receding sides contain flux in a circular or
multiple--component pattern. Closed contours are evident on both the
receding and approaching sides, and the galaxy exhibits a falling
$V_{rot}\sin i (r)$, indicating a warp. The line profile of the
galaxy, however, is quite symmetric, hiding the inherent dynamic
nonaxisymmetry, in agreement with HHMRvZ.

\subsubsection{UGC 6429}

As measured by KHL, UGC~6429 (Figure \ref{u6429}) is optically an
asymmetric Sc disk exhibiting two inner and two outer spiral arms,
with the southern arm terminating in a large, bright \ion{H}{2}
region. The \ion{H}{1} distribution exhibits an asymmetric ring
following the inner spiral arms, and a large high--column density
feature to the east, just outside $R_{25}$. This feature is an
extension of the northern optical outer spiral arm, which gives the
entire distribution a very asymmetric qualitative appearance.

Dynamically, the most apparent departures from axisymmetry in UGC~6429
are the closed contours on the approaching side of the velocity field,
and the change in position angle with radius to the north. There is no
dynamical feature corresponding to the eastern morphological
feature. The asymmetry in the integrated line profile of this galaxy
is due to the nonuniform weights given in integrating the line profile
due to the eastern density enhancement. The nearest neighbor listed in
the AGC to UGC~6429 is NGC~3668, 17$^\prime$ (371 kpc in projected
distance) to the southeast, at 3508~km~s$^{-1}$.

\subsubsection{NGC 4688}

NGC~4688 (Figure \ref{n4688}) is an outlying member of the Virgo
cluster, and is a member of LGG~292 (Garcia 1993). Its nearest
neighbor listed in that group is UGC~7983, an irregular galaxy
$43^\prime$ (213~kpc at the adopted distance of 17~Mpc) to the
southeast. NGC~4688 is a very asymmetric Scd galaxy in the optical,
where flocculent spiral arms extend away from the main body of the
galaxy to the northeast. Besides the tidally deformed galaxy NGC~5474,
NGC~4688 is the most optically asymmetric galaxy in the sample. The
\ion{H}{1} morphology, too, exhibits strong asymmetries, particularly
the very high--column density regions to the east--southeast just at
$R_{25}$. These regions correspond to the southernmost of the
flocculent optical arms as seen in {\it R\/}--band. Spiral structure,
in fact, can be seen throughout the \ion{H}{1} disk, corresponding to
flocculent structures in the optical galaxy. The low--column density
\ion{H}{1} is distributed preferentially to the northeast, lending an
even more qualitative lopsided appearance to the galaxy. This feature
contributes to a larger value of $R_{HI}/R_{25}$ to the north than to
the south. Quantitatively, as calculated in the next section, the
morphological asymmetry of the \ion{H}{1} distribution makes NGC~4688
the most morphologically asymmetric galaxy in gas as well as in
optical light.

Dynamically, too, NGC~4688 is disturbed. The receding side of the
velocity field exhibits closed contours which reopen again at
$1.5R_{25}$, a feature similar to that found in NGC~991 and
NGC~5474. This feature leads to a large asymmetry in the rotation
curve, which manifests itself as a large difference between observed
velocity on the approaching and receding sides at the same radius. The
global profile is unusual too, in that it exhibits very gradual
declines from the peaks to the noise level on both sides of the
profile.

A possible explanation for this very asymmetric behavior lies
6\farcm7 (33~kpc) to the northeast of the dynamical center of
NGC~4688: a small $3.2\times 10^7 M_\odot$ companion detected in the
\ion{H}{1} map. The peak of the \ion{H}{1} distribution of the
companion is located at B1950.0 coordinates $\alpha = 12^h45^m26^s\!\!.5$,
$\delta = +04^\circ42^\prime10^{\prime\prime}$. The systemic velocity
of the peak of the companion's emission is 991~km~s$^{-1}$.

The column density map of the companion is presented in Figure
\ref{n4688comp}. The top left frame of this figure presents the column
density of NGC~4688 and the companion for scaling; the top and bottom
right frames detail the companion itself.  The \ion{H}{1} column
density contours in the top right panel are overlaid on the Palomar
Observatory Sky Survey\footnote{The National Geographic Society --
Palomar Observatory Sky Atlas (POSS-I) was made by the California
Institute of Technology with grants from the National Geographic
Society.}  plate for this region digitized by the Space Telescope
Science Institute Digitized Sky Surveys\footnote{The Digitized Sky
Surveys were produced at the Space Telescope Science Institute under
U.S. Government grant NAG W--2166. The images of these surveys are
based on photographic data obtained using the Oschin Schmidt Telescope
on Palomar Mountain and the UK Schmidt Telescope. The plates were
processed into the present compressed digital form with the permission
of these institutions.}. The lower right panel depicts the POSS plate
alone. The companion is just resolved by our observations in the CS
array, and the maximum column density observed is $6\times
10^{20}$~cm$^{-2}$.  An optical counterpart to the \ion{H}{1} emission
is clearly seen. The observed location of the \ion{H}{1} emission,
however, is offset 20\arcsec\ to the southwest of the maximum of the
optical emission, but does correspond to a smaller local maximum. We
identify these two optical maxima as the galaxy pair
\hbox{CGCG~043--029.} The bottom left frame of Figure \ref{n4688comp}
illustrates the synthesised line profile of this object. The maximum
flux occurs in the 991~km~s$^{-1}$ channel. The gradual slope on the
receding side of the profile, however, suggests that there is probably
significant additional flux beyond our velocity limit of
1040~km~s$^{-1}$.

An \ion{H}{1} detection from the northeastern component of
\hbox{CGCG~043--029} appears in a table of \ion{H}{1} observations of faint
galaxies conducted with the Arecibo circular feed compiled by Lu {\it
et al.\/} (1993)\markcite{lu93}, except that the heliocentric velocity
reported there is 4889~km~s$^{-1}$. It is a possibility, therefore, that
the pair is an optical pair only, with the 4889~km~s$^{-1}$ emission
associated with the northeastern component, and the 991~km~s$^{-1}$
emission associated with the southwestern component. The data,
however, are too incomplete for us to make any supportable claims
regarding this object beyond its mere detection.

During the initial analysis of the 21~cm line data, we imaged the data
cube using a tapered beam of $40\times 45$ arcseconds size to examine
the data for evidence of an \ion{H}{1} bridge between NGC~4688 and
\hbox{CGCG~043--029.} No such bridge was detected.

\subsubsection{NGC 5474\label{N5474}}

NGC~5474 (Figure \ref{n5474}) is a peculiar spiral galaxy
approximately 44$^\prime$ (90~kpc at the adopted distance of 7~Mpc) to
the south of M101, and is tidally deformed into a very asymmetric and
disturbed morphology as observed by KHL in the optical.

NGC~5474 was first observed in \ion{H}{1} synthesis by van der Hulst
\& Huchtmeier (1979)\markcite{hulst5474} at the WSRT, with significant
\ion{H}{1} flux appearing in four channels with a velocity resolution of
27.2~km~s$^{-1}$. Subsequently, this galaxy was observed by RDH with
the D array at the VLA. Both groups indicate that the kinematic major
axis of this galaxy does not appear to form a straight line, but
rather is contorted into an ``S--''shape, and also note that the
kinematic center and morphological centroid of the \ion{H}{1}
distribution are approximately co--located with the center of the
outer optical isophotes, and not with the offset optical nucleus.
Hence, the true center of the galaxy is not the bright
``pseudonuclear'' region located to the north of the overall optical
light distribution. 

As mentioned in \S \ref{betterdata}, we are able to trace \ion{H}{1}
to a radius of approximately 400$^{\prime\prime}$, in contrast to the
250$^{\prime\prime}$ radius recovered by RDH and by van der Hulst \&
Huchtmeier. In addition to the ``clumps'' of neutral hydrogen found in
the high HI column density regions by RDH, we find a smooth HI
distribution beyond twice the optical radius. Even so, only half of
the reported single--dish flux for this galaxy was recovered, because
NGC~5474 and M101 share a common \ion{H}{1} distribution (Huchtmeier
\& Witzel 1979)\markcite{hw79}, and even the short spacings of the
D array are unable to recover the diffuse gas which envelopes
this galaxy and its dominant companion. We do, however, detect 50\%
more flux than that found by RDH.

Dynamically, the galaxy exhibits a number of bizarre features. RDH
detect 50$^\circ$ of position angle change from the dynamical center
to a radius of 250$^{\prime\prime}$. Our reanalysis agrees with their
analysis and also reveals the continuation of this trend further out,
where the position angle varies by a total of 95$^\circ$ from the
dynamical center to a 400$^{\prime\prime}$ radius. Position angles for
the approaching and receding sides also diverge in the outer regions,
with the difference in position angle from one side to the other of
$\approx 50^\circ$, although the greatest divergence occurs in a
region of small signal--to--noise (slightly below $3\sigma$) and may
therefore be an artifact of the model. These large position angle
variations as a function of radius are largely due to the very unusual
dynamic feature found 40$^{\prime\prime}$ to the south of the optical center
of the galaxy disk, in which the velocity contours fully reverse their
directions, a continuation of the ``S--''shape reported by the
previous observers. This feature also displaces the dynamical center
of the outer regions about 1$^\prime$ to the southwest of the optical
pseudonucleus center of light. The feature contributes to the
measurable asymmetry in the integrated single--dish line profile. Such
a feature may be the result a very warped disk viewed in a direction
which bisects the angle formed by the flat and warped portions of the
disk. The pseudonucleus itself has no discernible effect on the
dynamics of the disk.

RDH use a model in which the galaxy inclination is fixed at 21$^\circ$
to determine their rotation curve; if we fix the inclination at this
value, we recover the same curve. Both analyses agree on the general
shape of the curve, which rises out to about 100$\arcsec$, steeper on
the receding side, and flattens out with a shallow minimum at
200$\arcsec$.

\subsubsection{NGC 5701}

The barred Sa galaxy NGC~5701 (Figure \ref{n5701}) exhibits prominent
inner and outer ring structures in the optical image along with a
large bulge and bar at position angle $-5^\circ$. The optical galaxy
is embedded in a very large \ion{H}{1} envelope of $D_{HI}/D_{25} =
2.6$. More \ion{H}{1} flux is observed on the west side of the galaxy
than on the east by a factor of two, and the western drop--off is more
precipitous than the east. This feature is similar to that found in
NGC~1042 (\S~\ref{n1042}), where it is interpreted as a possible
remnant of tidal interaction. In this case, however, no companion to
NGC~5701 is seen in our optical or \ion{H}{1} data. It is, however,
identified by Garcia (1993)\markcite{garcia} as a member of the group
LGG~386, the nearest member of which to NGC~5701 is UGC~9385,
57\arcmin\ (431~kpc at the adopted distance of 26~Mpc) to the west.

In this connection, it is important to note that NGC~5701 was also
mapped at Arecibo by DuPrie \& Schneider (1996)\markcite{ds}. We agree
with their Figure 3, that the \ion{H}{1} is extended to $\sim
10$\arcmin; however, we do not recover the \ion{H}{1} ``cloud'' they
report 15\arcmin\ northwest of the center of NGC~5701 and of flux
5.5~Jy~km~s$^{-1}$. We do detect two flux peaks $4\sigma$ above
the noise 13\arcmin\ to the northwest in two channels (1495 and
1505~km~s$^{-1}$) in the archive data. However, because the emission
peaks occur in two different spatial locations separated by about a
beam--width, these peaks were excluded from the moment maps by the
windowing function. That these peaks occur at the same heliocentric
velocity as the ``cloud'' reported by DuPrie \& Schneider, and would
represent a total flux of 4.6~Jy~km~s$^{-1}$ suggests that the
``cloud'' is marginally detected in the synthesis data.

Nonlinearity of the velocity contours along the minor axis of NGC~5701
can be observed in the velocity field. This results in a dynamical
position angle change of 43$^\circ$ as a function of radius, similar
to the value obtained from analysis of NGC~1042. NGC~5701 also
exhibits significant differences between the values of $V_{rot}\sin i
(r)$ on either branch of the major axis, again similar to the rotation
curve of NGC~1042. NGC~5701 differs, however, in that it exhibits a
deep trough in $V_{rot}\sin i (r)$ at a radius of 210$\arcsec$. This
is suggestive of either a warp or strong streaming motions in this
galaxy. The synthesised line profile of NGC~5701 is consistent with
the single--dish profile presented by HHMRvZ. In agreement with them,
we find that the profile is symmetric.

\subsubsection{UGC 12732}

UGC~12732 (Figure \ref{u12732}), first observed by Schulman {\it et
al.\/} (1997)\markcite{schul}, is a small Sm galaxy with a roughly
symmetric optical appearance. Compared to the optical galaxy, the
neutral gas structure detected is extremely large: $D_{HI}/D_{25} =
3.1$ as given in Table \ref{globalHI}, and $M_{HI}/L_B$ is an order of
magnitude greater than that of the other targets. Indeed, UGC~12732 is
more gas--rich (or star--poor) than most Sm galaxies; $M_H/L_B$ of 3.2
is well in excess of the median of 0.78 for Sm and Im galaxies given
by Roberts \& Haynes (1994)\markcite{rh94}. The nearest neighbor of
known redshift within 1000~km~s$^{-1}$ to that of UGC~12732 is
NGC~7741, 45$^\prime$ (196 kpc) to the southeast, at 744~km~s$^{-1}$.

The contours in Figure \ref{u12732} were chosen to emphasize the
spiral structure slightly to the north of the optical galaxy. This
spiral structure does not appear to be associated with any optical
spiral structure. The region of maximum \ion{H}{1} column density is
located approximately 2$\arcmin$ to the SSE of the optical center,
which corresponds to the termination of the eastern spiral arm as seen
in the {\it R\/}--band. The outer \ion{H}{1} density contours are
centered on the optical center of light, as is a 1$\arcmin$ radius
ring surrounding the inner regions of the optical galaxy. The spiral
structures extend from this ring and are especially apparent to the
northwest. These structures contribute to a very high value of
morphological asymmetry for the neutral hydrogen distribution.  In
dynamics, a 30$^\circ$ position angle change on both the approaching
and receding sides of the galaxy, rising $V_{rot}\sin i (r)$, and an
asymmetric line profile contribute to overall dynamical nonaxisymmetry
in UGC~12732.

As Schulman {\it et al.\/} commented, the large $D_{HI}/D_{25}$ ratio
may be an indicator of inefficient star formation in this galaxy. The
$V_{rot}\sin i (r)$ curve and position angle fits we derive with the
tilted ring model differ from those derived by Schulman {\it et
al.\/}, in that the qualitative shapes of the curves significantly
differ between the two analyses. While in the Schulman {\it et al.\/}
analysis, the inclination was allowed to vary to obtain a flatter
rotation curve, our analysis does not attempt to disentangle $i$ from
$V_{rot}$, resulting in a rising $V_{rot}\sin i (r)$ curve and a
greater change in position angle as a function of radius.

\section{Comparing Morphological and Kinematic Asymmetry \label{correlations}}

\subsection{Obtaining the Various Measures of Asymmetry in a Galaxy}

We examined our data sets for any correlations between asymmetric
morphology and nonaxisymmetric dynamics. In doing so, we wished to search
for methods of quantifying deviations from kinematic axisymmetry,
including both non--planar and non--circular motions, which could then
be compared to the morphological measures. Dynamic models of the
galaxies allowed us to define several kinematic quantities which, we
hypothesised, might be correlated with asymmetric morphology. The
features we examined were: morphological asymmetry in {\it R\/}--band
and \ion{H}{1}, change of kinematic position angle with radius, change
of kinematic position angle with azimuth, and asymmetry in rotation
curves. Single--dish \ion{H}{1} line profile asymmetry was also
computed for these galaxies using the method described by HHMRvZ.

The zeroth order \ion{H}{1} moment maps were analyzed for
morphological asymmetry using the ``Method of Sectors'' described in
KHL. Each galaxy was divided into 8 equal--area sectors with the center
of the pattern located at the optical center of light, and the
position angle of the sector pattern oriented along the dynamical
position angle of the galaxy. The total \ion{H}{1} flux was measured
in each of the sectors. The maximum flux differences were then
normalized by the total integrated \ion{H}{1} flux through the
pattern. The result is the asymmetry parameter $A_f$:
\begin{equation}
A_f \equiv \frac{\Delta f^{max}}{\sum_n f_n}
\end{equation}
where $\Delta f^{max}$ is the largest flux difference between any two
sectors, and $f_n$ is the flux in the $n$th sector. 

{\it R\/}--band values of $A_{f,R}$ are presented here adapted from the
data used by KHL. In KHL, sector analysis of optical data was
presented by taking ratios of the maximum flux in a sector to the
minimum flux in a sector, and converting this to a magnitude
difference in the standard way. Thus, KHL present the data in terms of
magnitude differences $\Delta M$ between sectors rather than
normalized flux differences. There, a galaxy was said to be asymmetric
if a value of $\Delta M$ differed from zero by more than the $5\sigma$
RMS noise level, regardless of the magnitude of $\Delta M$ itself. For
this reason, some galaxies with large values of $\Delta M$ were
classified as symmetric because the errors were also large. Similarly,
small values of $\Delta M$ with small errors could indicate asymmetric
galaxies. Of the nine galaxies considered here, UGC~3685, NGC~5701, and
UGC~12732 were classified as symmetric; all others were classified
as asymmetric. Here, we use the same flux data as were used by KHL, but
present them as normalized flux differences $A_f$ for both the
optical and \ion{H}{1} data because they are more useful in this form
for understanding the symmetry properties of galaxy morphology in an
absolute and directly comparable sense.

Deviations from axisymmetric dynamics of the targets was measured using
data obtained from the model fitting. Because a symmetric, flat
rotating disk should exhibit a constant position angle, we took
changes in position angle in the model as quantifiers of
nonaxisymmetry. The first of these is the maximum change in position angle
on one side of the galaxy as a function of radius,
$\Delta\Gamma(r)$. Note that in cases where the position angle changes
steadily with radius, $\Delta\Gamma(r)$ can depend on the maximum
radius for which we can construct a model. As applied to these
face--on galaxies, $\Delta\Gamma(r)$ is most sensitive to non--planar, 
as opposed to non--circular motions. The second of these is the
difference between position angles on approaching and receding sides
of the model at the same radius averaged over all radii,
$\Delta\Gamma(\theta)$. $\Delta\Gamma(\theta)$ is sensitive to
non--circular motions and asymmetric warps.

Extending the analysis of S3vA, asymmetry in $V_{rot}\sin i (r)$
curves was quantified in the following way. The curve was folded over
the kinematic center of the model. The total area between the two
curves was then measured and normalized by the average area under the
curve, to produce the measure of global rotation asymmetry $S_2$:
\begin{equation}
S_2 \equiv 10^2 \times \frac{\int \left|\left|V_{rot}(r)\sin i(r)\right| -
\left|V_{rot}(-r)\sin i(-r)\right|\right|\,dr}{\int \onehalf
\left[\left|V_{rot}(r)\sin i(r)\right| + \left|V_{rot}(-r)\sin i(-r)\right|\right]\,dr} .
\end{equation}
Clearly, for more inclined galaxies where $V_{rot}$ and $i$ can be fit 
separately, $S_2$ can be modified to measure asymmetry in $V_{rot}(r)$ 
alone. In our sample of face--on galaxies, $S_2$ is very sensitive to
non--planar motions. In more inclined samples, $S_2$ would be more
sensitive to non--circular motions.

Because several groups (for example, Tifft \& Cocke
1988\markcite{tc88}, Richter \& Sancisi 1994\markcite{rs94}, and
HHMRvZ) have studied the symmetry properties of 21-cm line profiles as
an indicator of dynamical asymmetry in a galaxy, here too we apply the
method of line profile asymmetry described by HHMRvZ. The quantity
defined there is:
\begin{equation}
A_{l/h} \equiv \frac{\int_{V_l}^{V_{\rm med}} S\,dV}{\int_{V_{\rm
med}}^{V_h} S\,dV}
\end{equation}
where $V_l$ and $V_h$ are the low and high velocity limits,
respectively, of signal detection above $3\sigma$, and $V_{\rm med}$
is the median between them. The characteristic lopsidedness of the
line profile is given simply by
\begin{equation}
A_n \equiv \cases{A_{l/h},&if $A_{l/h}>1$\cr
		  1/A_{l/h},&otherwise.\cr}
\end{equation}
The values of $A_n$ of 1.00 and 1.03 obtained in this work for
NGC~3596 and NGC~5701, respectively, are in agreement with the values
of $1/0.99=1.01$ obtained by HHMRvZ within errors and differences in
velocity resolution. HHMRvZ found that approximately 50\% of the line
profiles studied exhibited asymmetry parameters $A_n \ge 1.05$, which
they considered to be indicative of overall asymmetry.

\begin{deluxetable}{lllrrrrr}
\tablenum{4}
\tablewidth{0pt}
\tablecaption{Target Symmetry Properties\label{symtab}}
\tablehead{
\colhead{(1)}&
\colhead{(2)}&
\colhead{(3)}&
\colhead{(4)}&
\colhead{(5)}&
\colhead{(6)}&
\colhead{(7)}
\\[.2ex]
\colhead{Galaxy}&
\colhead{$A_{f,R}$}&
\colhead{$A_{f,HI}$}&
\colhead{$\Delta \Gamma(r)$}&
\colhead{$\Delta \Gamma(\theta)$}&
\colhead{$S_2$}&
\colhead{$A_n$}
\\[.2ex]
\colhead{}&
\colhead{}&
\colhead{}&
\colhead{(degrees)}&
\colhead{(degrees)}&
\colhead{}&
\colhead{}}
\startdata
NGC 991  & 0.0254 $\pm$ .005 & 0.0393 $\pm$ 0.002 & 11.9 $\pm$ 2.5 & 5.0 $\pm$ 0.7   & 9.9  & 1.12\cr 
NGC 1042 & 0.0259 $\pm$ .005 & 0.0260 $\pm$ 0.001 & 43.4 $\pm$ 3.9 & 5.1 $\pm$ 0.4   & 19   & 1.04\cr 
UGC 3685 & 0.0536 $\pm$ .02  & 0.0646 $\pm$ 0.001 & 18.6 $\pm$ 3.9 & 6.2 $\pm$ 0.3   & 15   & 1.09\cr 
NGC 3596 & 0.0212 $\pm$ .001 & 0.0491 $\pm$ 0.002 & 27.2 $\pm$ 4.0 & 6.6  $\pm$ 0.4  & 3.2  & 1.00\cr
UGC 6429 & 0.0189 $\pm$ .003 & 0.0607 $\pm$ 0.003 & 26.1 $\pm$ 5.0 & 6.4  $\pm$ 1.5  & 9.4  & 1.08\cr
NGC 4688 & 0.0725 $\pm$ .005 & 0.0690 $\pm$ 0.003 & 25.9 $\pm$ 3.2 & 9.0  $\pm$ 0.8  & 15   & 1.07\cr
NGC 5474 & $>0.1$            & 0.0576 $\pm$ 0.01  & 95.4 $\pm$ 6.4 & 14.9 $\pm$ 0.7  & 10   & 1.18\cr 
NGC 5701 & 0.0322 $\pm$ .01  & 0.0419 $\pm$ 0.008 & 43.0 $\pm$ 3.0 & 4.1  $\pm$ 0.3  & 3.6  & 1.03\cr 
UGC 12732& 0.0297 $\pm$ .02  & 0.0613 $\pm$ 0.009 & 15.3 $\pm$ 5.7 & 5.5  $\pm$ 1.4  & 6.0  & 1.28\cr 
\enddata
\end{deluxetable}

Results for the asymmetry measures for each target are presented in
Table \ref{symtab}.  A plot of $A_{f,R}$ versus $A_{f,HI}$ and the
plots of all morphological/dynamical pairs are presented in Figure
\ref{scatter}. This figure excludes points for NGC~5474, because of
its large lower bound on $A_{f,R}$. Errors on $A_n$ are the size of
the points, while systematic errors on $S_2$ we estimate at about
$20\%$. Little correlation is apparent in our data set between
\ion{H}{1} morphology and dynamics for disk galaxies in the
field. In our sample, there is a relationship between the \ion{H}{1}
and {\it R\/}--band morphologies in which large values of $A_{f,R}$
imply large values of $A_{f,HI}$, although not the converse. It should
not be surprising to find that optical and \ion{H}{1} morphology are
related, because \ion{H}{1} spiral structure tends to follow the
optical spiral structure.

All pairs of morphological versus dynamical parameters were also
examined. The only statistically significant relationship between
$A_f$ and any of the calculated dynamical parameters is a weak
relation with $\Delta\Gamma(\theta)$. Large values of
$\Delta\Gamma(\theta)$ imply large values of $A_{f,HI}$ within our
sample, but again, not the converse.  It is also interesting that some
of the most morphologically symmetric galaxies exhibit asymmetry in
their line profiles as measured by $A_n > 1.05$. This qualitative
observation and the lack of correlation between these parameters seems
to support the casual observations of HHMRvZ and others who have noted
a similar non--correlation. For instance, HHMRvZ comment that very
morphologically disturbed galaxies such as NGC~1637 often have
undisturbed \ion{H}{1} line profiles.

The morphology of the asymmetries themselves is also worthy of
study. KHL classified asymmetric morphology into three categories
according to the relative locations of the maximum and minimum flux
sectors in the 8--segment {\it R\/}--band analysis. The categories
were:: ``bisymmetric,'' in which the extreme sectors were adjacent,
``boxy,'' in which the extreme sectors were 90$^\circ$ apart, and
``lopsided,'' in which the extreme sectors were 135$^\circ$ or
180$^\circ$ apart. Of the galaxies in this study classified as
asymmetric by KHL, NGC~991, UGC~6429, and NGC~5474 were classified as
lopsided, while NGC~1042, NGC~3596, and NGC~4688 were classified as
boxy. We find that in the \ion{H}{1}, all of these galaxies exhibit
lopsided asymmetry, except for NGC~1042. NGC~1042 exhibits boxy
asymmetry, caused by the peculiar perpendicularity of its
morphological and dynamical major axes. The remaining galaxies in the
current sample were classified as symmetric by KHL, and therefore were
not categorized in this way in {\it R\/}--band. Of these galaxies,
NGC~5701 and UGC~3685 are boxy in the \ion{H}{1}, while UGC~12732 is
lopsided. None of the observed galaxies were ``bisymmetric'' in the
\ion{H}{1}.

It is also of interest to search for asymmetry trends as a function of
global gas parameters. Figure \ref{scatter2} illustrates the
relationships between $A_{f,HI}$ and and the relevant global
\ion{H}{1} parameters of the targets as given in Table
\ref{globalHI}. Again, no clear relationships are apparent. The
relationship between dynamical asymmetry and global \ion{H}{1}
properties is even more tenuous. Figure \ref{scatterx} illustrates
this clearly. There are no detectable dependencies of any global
property of the gas on the rotation asymmetry $S_2$.

\subsection{Comparison of Current Techniques} 

As discussed in \S\ref{intro}, several studies of optical
morphological asymmetry have been conducted, in general using a number
of different quantitative measures. One of these, Conselice {\it et
al.\/} (1999)\markcite{C99}, includes NGC~3596 and NGC~5701 in its
galaxy sample. They employ a method in which a galaxy image is rotated
through some angle $\phi$ and then subtracted from the original,
unrotated image. The normalized residuals are taken as the asymmetry
measurement. Our measure of {\it R-\/} band morphology is in
disagreement with Conselice {\it et al.\/}, insofar as NGC~3596 is one
of our most symmetric galaxies as measured by $A_{f,R}$, and NGC~5701
is measured to have moderate asymmetry by the same quantifier. In
contrast, Conselice {\it et al.\/} find that NGC~3596 is more
asymmetric than about 75\% of their sample, and NGC~5701 is one of the
most symmetric. These differences are likely due to several factors;
primarily, our methods are fundamentally different, and measure
different types of asymmetry. In addition, centering issues are very
likely to be an important factor in the disagreement here, as
Conselice {\it et al.\/} use a different centering method from that
used here (see below).

The dynamical symmetry properties which we have derived represent a
wide range of dynamical variables, such that the combination of these
parameters allows for the study of the global properties of a galaxy
which are the most easily accessible to the casual observer. Other
techniques for the determination of dynamic asymmetry have also been
proposed, such as that by SFdZ\markcite{SFdZ}. These authors present
methods by which lopsided spiral modes and global elongation in the
galactic potential may be inferred from the presence of
nonaxisymmetric motions in the observed velocity field. In light of
theoretical models such as that of Jog (1997)\markcite{jog}, it is
clear that perturbations in the gravitational potential can manifest
as perturbations in the observed velocity fields.

Unlike the work presented here, determining the spiral structure of
the potentials of disk galaxies using the method of SFdZ leads
immediately to information about the total mass distribution in
galactic halos. Nevertheless, the method is complex and is most
applicable over a narrow range of inclinations. The parameters
calculated in the present work are calculable over a wide range of
galaxy inclinations and are quickly attainable from basic data, and
thus are better suited to quickly identifying asymmetric galaxies from
large samples. Once identified, more detailed methods would be
invaluable for deriving the perturbed galactic potential of individual
galaxies.

A method based on SFdZ was used by S3vA to quantify the asymmetry in
the two galaxies presented in that paper, NGC~4395 and DDO~9. There,
they find that both galaxies, selected for strong kinematic
lopsidedness, exhibit asymmetric rotation curves. To test this
observation, we attempted to apply the SFdZ method to our sample of
nine galaxies to find the zeroth through third harmonic components of
each galaxy's velocity field as a function of radius. However, because 
of the low inclinations of these galaxies, satisfactory decomposition
into harmonic kinematic components for the comparison was impossible,
as measured in part by consistently large values of the third cosine
coefficient, $\hat c_3$, which should be zero when the inclination of
the tilted ring fit is correct.

Closely connected with both the present method and those of Conselice
{\it et al.\/}, SFdZ and others is the question of centering. Although
not yet utilized as a parameter of asymmetry itself, the question of
finding the kinematic center and/or the center of light of galaxy is
critical to the question of galactic symmetry; imprecise determination
of the center of the galaxy can easily introduce unacceptable errors
in the determination of symmetry properties. In the present method and
that of Conselice {\it et al.\/}, improper centering creates a false
dipole in the overlay pattern (or the rotation) for determining
morphology, as it does for true Fourier morphology techniques such as
that of Zaritsky \& Rix (1997)\markcite{zr97}. Improper centering also
contaminates measures of $S_2$ and $\Delta\Gamma$. In this work,
centers for kinematic parameters were chosen iteratively for best
chi--square fit in the tilted ring model, while the centers for the
morphological parameters were chosen as the centers of light of the
optical galaxy. Conselice {\it et al.\/} iteratively find the
centering which minimizes their morphological asymmetry
parameter. SFdZ recommend selecting centers which minimize the total
residual velocity.

Beauvais \& Bothun (1999)\markcite{bb99} also consider the issue of
centering when applying a tilted--ring model to Fabry--Perot
data. They use three methods for centering of the galaxy: photometric
ellipse fitting, finding the center of symmetry of the tilted--ring
model, and fitting the center as a free parameter in the tilted--ring
model. They report that each of these centering methods yielded
equally good model velocity fields, but that it was common for the
center to vary considerably in the outer radii of the galaxy when
allowed to vary freely, and interpret this as possibly indicating
actual misalignment of the inner and outer rotation axes.

Because centering is so critical, nearly all studies of asymmetry must
choose one or two methods for finding the center of a galaxy. These
derived centers themselves can also be used as indicators of
lopsidedness. Offsets of the kinematic center from the optical center
of light and dependence on the derived kinematic center on radius can
both be indicators of lopsided dynamics. In this work, the sample
galaxies were chosen on the basis of their face--on inclinations for
the purpose of examining the optical morphology, making the kinematic
centers not sufficiently certain for comparison with morphological
centers. However, future work with more inclined galaxies may be able
to use centering itself as a symmetry indicator.

\section{Summary \label{conclusions}}

This work represents a concurrent analysis of both optical and 21-cm
symmetry properties of a sample of disk galaxies. We have presented
methods by which the symmetry properties of a galaxy can be quickly
computed over a wide range of dynamic variables. These parameters are
sensitive to both non-circular and non-planar motions, depending in
large part on the inclinations of the galaxies. In this case, the
face--on inclinations of the targets make the parameters most
sensitive to non--planar motions.

In our sample presented here, tidal interaction with companion objects
dominates the asymmetry in NGC~5474 and cannot be ruled out for
NGC~1042 and NGC~4688. Nevertheless, dynamical nonaxisymmetries
clearly prosper in more isolated and even morphologically symmetric
galaxies such as UGC~12732 and UGC~3685 as well. Pervasive deviations
from axisymmetry even outside of tidal interactions seem to indicate
that asymmetric galaxies are stable and long--lived phenomena. Our
data seem, however, to indicate that there is little or no connection
between lopsided morphology and non--planar motions in such
galaxies. This lack of clear connection between morphology and
warping seems to imply that separate physical mechanisms may be at
work in each domain to produce deviations from flat, axisymmetric disks.

The methods presented in this paper can be quickly applied to large
samples of galaxies to identify deviations from axisymmetry in
dynamics and morphology. The small sample size of this initial study
and the kinematic uncertainties produced by the sample galaxies'
face--on inclinations can only rule out the strongest of correlations
among the various parameters. Larger samples and samples at varying
inclinations are required to truly study the relationships between
these parameters at more than a cursory level. A future study with
galaxies of a wider range of inclinations will also allow for
additional parameters such as kinematic offsets which can then be
adequately studied. Future numerical work can also benefit from an
easily applied method to track and identify departures from
symmetry. Such numerical work could, for instance, study the origins
and longevity of asymmetries, and predict the course of their
evolution in spiral galaxies.

\acknowledgements
This research has been partially supported by NSF grants AST--9528860
and AST--9900695 to MPH, and by National Space Grant College and
Fellowship Program grant NGT--40019 to DAK. We also wish to
acknowledge the observers of the archived data sets for NGC~5474,
NGC~5701, and UGC~12732 for making the archived observations, and to
the VLA data archive for making these data available. This research
has made use of the NASA/IPAC Extragalactic Database (NED) which is
operated by the Jet Propulsion Laboratory, California Institute of
Technology, under contract with the National Aeronautics and Space
Administration, and of data obtained through the High Energy
Astrophysics Science Archive Research Center Online Service, provided
by the NASA/Goddard Space Flight Center.

\eject

\begin{figure}

\caption[]{Channel maps for each of the six new data sets in the
sample: (a)~NGC~991; (b)~NGC~1042; (c)~UGC~3685; (d)~NGC~3596;
(e)~UGC~6429; (f)~NGC~3688. Channel velocity is given in each
sub-frame, and contours are at the (-2, 2, 3, 6, 9, $\ldots$)$\sigma$
levels, where $\sigma$ is as given in Table \ref{globalHI}. Grayscales
are at the (6, 12, 18, $\ldots$)$\sigma$ levels.\label{chanmaps}}
\end{figure}

\begin{figure}
\caption[]{VLA \ion{H}{1} synthesis results for NGC~991. Top left panel:
\ion{H}{1} column density contours detected at the VLA, with light
gray contour at $3.4\times 10^{19}$~cm$^{-2}$ and solid contours at
$\{10.2, 23.8, 37.4, 51.0,\ldots\}\times 10^{19}$~cm$^{-2}$, overlaid on
the {\it R\/}--band image obtained by KHL. The first solid contour
represents the $3\sigma$ noise level. Top right panel: The same
\ion{H}{1} column density contours as in the top left panel, plotted
over a linear grayscale column density map. The circle is of diameter
$D_{25}$ and is centered on the {\it R\/}--band center of light of the
galaxy. Center left panel: The position-velocity diagram obtained from
the \ion{H}{1} data cube. The slice was made along the dynamical major
axis $\Gamma = 295^\circ$ determined by averaging the dynamical
position angle over radii $r < R_{25}$. The $x$--axis indicates
arcminutes northwest of the dynamical center, and the dotted vertical
lines identify $R_{25}$. Center right panel: The \ion{H}{1} observed
velocity field with contours and grayscales from 1490~km~s$^{-1}$
(east, light shading) to 1570~km~s$^{-1}$ (west, dark shading) in
increments of 10~km~s$^{-1}$. The 1500, 1540, and 1580~km~s$^{-1}$
contours are identified in the panel. The circle is of diameter
$D_{25}$ and is centered on the {\it R\/}--band center of light of the
galaxy. Bottom left panel: The synthesized \ion{H}{1} line profile
obtained by integrating the data cube along both spatial axes. Bottom
right panel: The rotation $V_{rot}\sin i$ and dynamical major axis
position angle $\Gamma$ as a function of radius in arcseconds,
obtained from the tilted--ring model. The data for the approaching
side (circles; solid lines) and the receding side (triangles; dashed
lines) are presented superposed. \label{n991} }
\end{figure}

\begin{figure}
\caption[]{VLA \ion{H}{1} synthesis results for NGC~1042. Top left panel:
\ion{H}{1} column density contours detected at the VLA, with light
gray contour at $2.6\times 10^{19}$~cm$^{-2}$ and solid contours at
$\{7.8, 23.4, 39.0, 54.6,\ldots\}\times 10^{19}$~cm$^{-2}$, overlaid on
the {\it R\/}--band image obtained by KHL. The first solid contour
represents the $3\sigma$ noise level. Top right panel: The same
\ion{H}{1} column density contours as in the top left panel, plotted
over a linear grayscale column density map. The circle is of diameter
$D_{25}$ and is centered on the {\it R\/}--band center of light of the
galaxy. Center left panel: The position-velocity diagram obtained from
the \ion{H}{1} data cube. The slice was made along the dynamical major
axis $\Gamma = 292^\circ$ determined by averaging the dynamical
position angle over radii $r < R_{25}$. The $x$--axis indicates
arcminutes northwest of the dynamical center, and the dotted vertical
lines identify $R_{25}$. Center right panel: The \ion{H}{1} observed
velocity field with contours and grayscales from 1320~km~s$^{-1}$
(east, light shading) to 1420~km~s$^{-1}$ (west, dark shading) in
increments of 10~km~s$^{-1}$. The 1330, 1370, and 1410~km~s$^{-1}$
contours are identified in the panel. The circle is of diameter
$D_{25}$ and is centered on the {\it R\/}--band center of light of the
galaxy. Bottom left panel: The synthesized \ion{H}{1} line profile
obtained by integrating the data cube along both spatial axes. Bottom
right panel: The rotation $V_{rot}\sin i$ and dynamical major axis
position angle $\Gamma$ as a function of radius in arcseconds,
obtained from the tilted--ring model. The data for the approaching
side (circles; solid lines) and the receding side (triangles; dashed
lines) are presented superposed. \label{n1042fig} }
\end{figure}

\begin{figure}
\caption[]{VLA \ion{H}{1} synthesis results for UGC~3685. Top left panel:
\ion{H}{1} column density contours detected at the VLA, with light
gray contour at $1.1\times 10^{19}$~cm$^{-2}$ and solid contours at
$\{3.3, 13.2, 23.1, 33.0,\ldots\}\times 10^{19}$~cm$^{-2}$, overlaid on
the {\it R\/}--band image obtained by KHL. The first solid contour
represents the $3\sigma$ noise level. Top right panel: The same
\ion{H}{1} column density contours as in the top left panel, plotted
over a linear grayscale column density map. The circle is of diameter
$D_{25}$ and is centered on the {\it R\/}--band center of light of the
galaxy. Center left panel: The position-velocity diagram obtained from
the \ion{H}{1} data cube. The slice was made along the dynamical major
axis $\Gamma = 299^\circ$ determined by averaging the dynamical
position angle over radii $r < R_{25}$. The $x$--axis indicates
arcminutes northwest of the dynamical center, and the dotted vertical
lines identify $R_{25}$. Center right panel: The \ion{H}{1} observed
velocity field with contours and grayscales from 1730~km~s$^{-1}$
(east, light shading) to 1830~km~s$^{-1}$ (west, dark shading) in
increments of 10~km~s$^{-1}$. The 1780, 1810, and 1830~km~s$^{-1}$
contours are identified in the panel. The circle is of diameter
$D_{25}$ and is centered on the {\it R\/}--band center of light of the
galaxy. Bottom left panel: The synthesized \ion{H}{1} line profile
obtained by integrating the data cube along both spatial axes. Bottom
right panel: The rotation $V_{rot}\sin i$ and dynamical major axis
position angle $\Gamma$ as a function of radius in arcseconds,
obtained from the tilted--ring model. The data for the approaching
side (circles; solid lines) and the receding side (triangles; dashed
lines) are presented superposed. \label{u3685} }
\end{figure}

\begin{figure}
\caption[]{VLA \ion{H}{1} synthesis results for NGC~3596. Top left panel:
\ion{H}{1} column density contours detected at the VLA, with contours
at $\{4.05, 16.2, 28.35, 40.5,\ldots\}\times 10^{19}$~cm$^{-2}$,
overlaid on the {\it R\/}--band image obtained by KHL. The first
contour represents the $3\sigma$ noise level. Top right panel: The
same \ion{H}{1} column density contours as in the top left panel,
plotted over a linear grayscale column density map. The circle is of
diameter $D_{25}$ and is centered on the {\it R\/}--band center of
light of the galaxy. Center left panel: The position-velocity diagram
obtained from the \ion{H}{1} data cube. The slice was made along the
dynamical major axis $\Gamma = 80^\circ$ determined by averaging the
dynamical position angle over radii $r < R_{25}$. The $x$--axis
indicates arcminutes west of the dynamical center, and the dotted
vertical lines identify $R_{25}$. Center right panel: The \ion{H}{1}
observed velocity field with contours and grayscales from
1130~km~s$^{-1}$ (west, light shading) to 1260~km~s$^{-1}$ (east, dark
shading) in increments of 10~km~s$^{-1}$. The 1170, 1200, and
1220~km~s$^{-1}$ contours are identified in the panel. The circle is
of diameter $D_{25}$ and is centered on the {\it R\/}--band center of
light of the galaxy. Bottom left panel: The synthesized \ion{H}{1}
line profile obtained by integrating the data cube along both spatial
axes. Bottom right panel: The rotation $V_{rot}\sin i$ and dynamical
major axis position angle $\Gamma$ as a function of radius in
arcseconds, obtained from the tilted--ring model. The data for the
approaching side (circles; solid lines) and the receding side
(triangles; dashed lines) are presented superposed. \label{n3596} }
\end{figure}

\begin{figure}
\caption[]{VLA \ion{H}{1} synthesis results for UGC~6429. Top left panel:
\ion{H}{1} column density contours detected at the VLA, with contours
at $\{5.4, 16.2, 27, 37.8,\ldots\}\times 10^{19}$~cm$^{-2}$, overlaid
on the {\it R\/}--band image obtained by KHL. The first
contour represents the $3\sigma$ noise level. Top right panel: The
same \ion{H}{1} column density contours as in the top left panel,
plotted over a linear grayscale column density map. The circle is of
diameter $D_{25}$ and is centered on the {\it R\/}--band center of
light of the galaxy. Center left panel: The position-velocity diagram
obtained from the \ion{H}{1} data cube. The slice was made along the
dynamical major axis $\Gamma = -9^\circ$ determined by averaging the
dynamical position angle over radii $r < R_{25}$. The $x$--axis
indicates arcminutes north of the dynamical center, and the dotted
vertical lines identify $R_{25}$. Center right panel: The \ion{H}{1}
observed velocity field with contours and grayscales from
3700~km~s$^{-1}$ (south, light shading) to 3770~km~s$^{-1}$ (north, dark
shading) in increments of 5~km~s$^{-1}$. The 3710, 3735, and
3750~km~s$^{-1}$ contours are identified in the panel. The circle is
of diameter $D_{25}$ and is centered on the {\it R\/}--band center of
light of the galaxy. Bottom left panel: The synthesized \ion{H}{1}
line profile obtained by integrating the data cube along both spatial
axes. Bottom right panel: The rotation $V_{rot}\sin i$ and dynamical
major axis position angle $\Gamma$ as a function of radius in
arcseconds, obtained from the tilted--ring model. The data for the
approaching side (circles; solid lines) and the receding side
(triangles; dashed lines) are presented superposed. \label{u6429} }
\end{figure}

\begin{figure}
\caption[]{VLA \ion{H}{1} synthesis results for NGC~4688. Top left panel:
\ion{H}{1} column density contours detected at the VLA, with contours
at $\{7.2, 21.6, 36.0, 50.4,\ldots\}\times 10^{19}$~cm$^{-2}$, overlaid
on the {\it R\/}--band image obtained by KHL. The first
contour represents the $3\sigma$ noise level. Top right panel: The
same \ion{H}{1} column density contours as in the top left panel,
plotted over a linear grayscale column density map. The circle is of
diameter $D_{25}$ and is centered on the {\it R\/}--band center of
light of the galaxy. Center left panel: The position-velocity diagram
obtained from the \ion{H}{1} data cube. The slice was made along the
dynamical major axis $\Gamma = 3^\circ$ determined by averaging the
dynamical position angle over radii $r < R_{25}$. The $x$--axis
indicates arcminutes north of the dynamical center, and the dotted
vertical lines identify $R_{25}$. Center right panel: The \ion{H}{1}
observed velocity field with contours and grayscales from
960~km~s$^{-1}$ (south, light shading) to 1010~km~s$^{-1}$ (north,
dark shading) in increments of 5~km~s$^{-1}$. The 960, 980, and
1000~km~s$^{-1}$ contours are identified in the panel. The circle is
of diameter $D_{25}$ and is centered on the {\it R\/}--band center of
light of the galaxy. Bottom left panel: The synthesized \ion{H}{1}
line profile obtained by integrating the data cube along both spatial
axes in the region containing flux from NGC~4688. The profile does not
include flux from the companion. Bottom right panel: The rotation
$V_{rot}\sin i$ and dynamical major axis position angle $\Gamma$ as a
function of radius in arcseconds, obtained from the tilted--ring
model. The data for the approaching side (circles; solid lines) and
the receding side (triangles; dashed lines) are presented
superposed. \label{n4688} }
\end{figure}

\begin{figure}
\caption[]{Wide field VLA \ion{H}{1} synthesis data for
\hbox{CGCG~043--029,} the companion of NGC~4688. Top left panel: The
\ion{H}{1} column density of NGC~4688 and companion. The contours and
grayscales are as in figure \ref{n4688}. Top right panel: The
\ion{H}{1} column density of the companion object alone, overlaid on
the POSS image of the region. Bottom left panel: The integrated line
profile of the companion alone, obtained by integrating both spatial
axes of the data cube in the region containing flux from the
companion. Bottom right panel: the POSS image of \hbox{CGCG~043--029}
without the \ion{H}{1} contours, showing clearly the two distinct
optical maxima. \label{n4688comp}}
\end{figure}

\begin{figure}
\caption[]{VLA \ion{H}{1} synthesis results for NGC~5474 obtained from
VLA archive data originally reported by RDH. Top left panel:
\ion{H}{1} column density contours detected at the VLA, with contours
at $\{1.8, 10.8, 19.8, 28.8,\ldots\}\times 10^{19}$~cm$^{-2}$,
overlaid on the {\it R\/}--band image obtained by KHL. The first
contour represents the $3\sigma$ noise level. Top right panel: The
same \ion{H}{1} column density contours as in the top left panel,
plotted over a linear grayscale column density map. The circle is of
diameter $D_{25}$ and is centered on the {\it R\/}--band center of
light of the galaxy. Center left panel: The synthesized \ion{H}{1}
line profile obtained by integrating the data cube along both spatial
axes. Center right panel: The \ion{H}{1} observed velocity field with
contours and grayscales from 259~km~s$^{-1}$ (west, light shading) to
297~km~s$^{-1}$ (east, dark shading) in increments of
4~km~s$^{-1}$. The 261, and 295~km~s$^{-1}$ contours are identified in
the panel. The circle is of diameter $D_{25}$ and is centered on the
dynamical center of the galaxy as determined by tilted--ring
fitting. Bottom left panel: The rotation $V_{rot}\sin i$ and dynamical
major axis position angle $\Gamma$ as a function of radius in
arcseconds, obtained from the tilted--ring model. The data for the
approaching side (circles; solid lines) and the receding side
(triangles; dashed lines) are presented superposed. \label{n5474} }
\end{figure}

\begin{figure}
\caption[]{VLA \ion{H}{1} synthesis results for NGC~5701 obtained from
VLA archive data originally observed by S.~E.~Schneider. Top left panel:
\ion{H}{1} column density contours detected at the VLA, with contours
at $\{2.9, 7.8, 12.7, 17.6,\ldots\}\times 10^{19}$~cm$^{-2}$, overlaid
on the {\it R\/}--band image obtained by KHL. The first contour
represents the $3\sigma$ noise level. Top right panel: The same
\ion{H}{1} column density contours as in the top left panel, plotted
over a linear grayscale column density map. The circle is of diameter
$D_{25}$ and is centered on the {\it R\/}--band center of light of the
galaxy. Center left panel: The synthesized \ion{H}{1} line profile
obtained by integrating the data cube along both spatial axes. Center
right panel: The \ion{H}{1} observed velocity field with contours and
grayscales from 1440~km~s$^{-1}$ (west, light shading) to
1560~km~s$^{-1}$ (east, dark shading) in increments of
10~km~s$^{-1}$. The 1450, 1470, and 1550~km~s$^{-1}$ contours are
identified in the panel. The circle is of diameter $D_{25}$ and is
centered on the {\it R\/}--band center of light of the galaxy. Bottom
left panel: The rotation $V_{rot}\sin i$ and dynamical major axis
position angle $\Gamma$ as a function of radius in arcseconds,
obtained from the tilted--ring model. The data for the approaching
side (circles; solid lines) and the receding side (triangles; dashed
lines) are presented superposed. \label{n5701} }
\end{figure}

\begin{figure}
\caption[]{VLA \ion{H}{1} synthesis results for UGC~12732 obtained from
VLA archive data originally observed by Schulman {\it et al.\/}
(1997). Top left panel: \ion{H}{1} column density contours detected at
the VLA, with light gray contour at $6.7\times 10^{18}$~cm$^{-2}$ and
solid contours at $\{2.0, 11.4, 20.8, 30.2,\ldots\}\times
10^{19}$~cm$^{-2}$, overlaid on the {\it R\/}--band image obtained by
KHL. The first solid contour represents the $3\sigma$ noise level. Top
right panel: The same \ion{H}{1} column density contours as in the top
left panel, plotted over a linear grayscale column density map. The
circle is of diameter $D_{25}$ and is centered on the {\it R\/}--band
center of light of the galaxy. Center left panel: The synthesized
\ion{H}{1} line profile obtained by integrating the data cube along
both spatial axes. Center right panel: The
\ion{H}{1} observed velocity field with contours and grayscales from
680~km~s$^{-1}$ (south, light shading) to 810~km~s$^{-1}$ (east, dark
shading) in increments of 5~km~s$^{-1}$. The 720, 760, and
810~km~s$^{-1}$ contours are identified in the panel. The circle is of
diameter $D_{25}$ and is centered on the {\it R\/}--band center of
light of the galaxy. Bottom left panel: The rotation $V_{rot}\sin i$
and dynamical major axis position angle $\Gamma$ as a function of
radius in arcseconds, obtained from the tilted--ring model. The data
for the approaching side (circles; solid lines) and the receding side
(triangles; dashed lines) are presented superposed. \label{u12732} }
\end{figure}

\begin{figure}
\caption[]{Illustrations of the observed relationships between the
morphological asymmetry parameters ($A_{f,HI}$ and $A_{f,R}$) and the
dynamical parameters measuring position angle change with radius
$\Delta\Gamma(r)$, position angle change from approaching to receding
side $\Delta\Gamma(\theta)$, $V_{rot}\sin i (r)$ curve asymmetry
$S_2$, and integrated \ion{H}{1} line profile asymmetry
$A_n$.\label{scatter}}
\end{figure}

\begin{figure}
\caption[]{Illustrations of the relationships between \ion{H}{1}
morphology and the global \ion{H}{1} parameters, from top to bottom,
mass-to-light ratio in solar units, \ion{H}{1} mass, \ion{H}{1}
diameter in terms of optical diameter $D_{25}$, and projected distance
to the nearest neighbor. The positions of UGC~3685 and NGC~3596, which
have no neighbors within 1\arcdeg are indicated by vertical
arrows.\label{scatter2}}
\end{figure}

\begin{figure}
\caption[]{Illustrations of the relationships between rotation 
curve asymmetry $S_2$ and the global \ion{H}{1} parameters examined in
Figure \ref{scatter2}.\label{scatterx}}
\end{figure}

\end{document}